\documentclass[prd,preprint,superscriptaddress,showpacs,byrevtex]{revtex4}
\usepackage{epsfig}

\newcommand{\be}{\begin{equation}}
\newcommand{\ee}{\end{equation}}
\newcommand{\bea}{\begin{eqnarray}}
\newcommand{\eea}{\end{eqnarray}}
\def\simgt{\rlap{\lower 3.5 pt \hbox{$\mathchar \sim$}} \raise 1pt
Ê \hbox {$>$}}

\begin{document}

\title{Top Quark Production from Black Holes at the CERN LHC}

\author{Andrew Chamblin}\email{jashb@herald.ox.ac.uk}
\affiliation{Department of Physics, University of Louisville,
Louisville, KY 49292, USA}

\author{Fred Cooper} \email{fcooper@nsf.gov}
\affiliation{ National Science Foundation, Division of Physics,
Arlington, VA 22230, USA; and Santa Fe Institute, Santa Fe, NM 87501, USA}

\author{Gouranga C. Nayak} \email{gnayak@uic.edu}
\affiliation{ Department of Physics, University of Illinois, Chicago, IL 60607 USA
}

\date{\today}

\begin{abstract}

LHC is expected to be a top quark factory. If the fundamental Planck scale is near a TeV,
then we also expect the top quarks to be produced from black holes via Hawking radiation.
In this paper we calculate the cross sections for top quark production from black holes at the LHC
and compare it with the direct top quark cross section via parton fusion processes at next-to-next-to-leading
order (NNLO). We find that the top quark production from black holes can be larger or smaller than
the pQCD predictions at NNLO depending upon the Planck mass and black hole mass.
Hence the observation of very high rates for massive particle production (top quarks,
higgs or supersymmetry) at the LHC may be an useful signature for black hole production.

\end{abstract}
\maketitle

\newpage

\section{ Introduction }

Andrew Chamblin was a very good friend and a much valued collaborator-
we greatly miss him. This paper was initiated by Andrew.

It is now generally accepted that the scale of quantum gravity {\it could be}
as low as one TeV \cite{folks}
and hence there can be graviton, radion and black hole production at LHC
\cite{ppbf,pp,pp1,pp2,pp3,ag,ppch,ppk,ppu,park,hof,more,gram,cham,pp4,pp5,pp6}.
If such processes occur then LHC collider experiments \cite{gravr,gravr1}
can probe TeV scale quantum gravity. One of the most exciting aspects of this
will be the production of black holes in particle accelerators. These
`brane-world' black holes will be our first window into the extra
dimensions of space predicted by string theory, and required by the
several brane-world scenarios that provide for a low energy
Planck scale \cite{large}. As the black hole
masses at the LHC are relatively small (3-7 TeV) and the temperatures of the
black holes are very high ($\sim$ 1 TeV) the black holes can be a source for
top quark production via Hawking radiation.
In fact there can be an enormous amount of heavy (supersymmetry and higgs) particle
production from black holes \cite{susy,higgs}, much more than expected from
normal pQCD processes.
This comes about from two competing effects as the Planck scale
increases: 1) top quark production from black holes increases
because the temperature of the black holes increases as the
Planck scale increases for fixed black hole masses
(see below) and 2) the cross section for black hole
production decreases \cite{gray4,bv,grayd,cooper}.
Reviews of this exciting field are given in \cite{reviews}.

In this paper we calculate top quark production cross sections
from TeV scale black hole at the LHC via Hawking radiation and compare them with the
direct pQCD parton fusion processes at next-to-next-to-leading order (NNLO).
We find that the top quark production cross sections from
black holes at the LHC can be larger or smaller than those
from pQCD processes at NNLO depending on the value of the TeV scale
Planck mass and the black hole masses. We find that as long as the
temperature of the black holes is of the order of TeV,
the top quark production cross section from the
black holes does not depend very much on the top quark mass $M_t$.
On the other hand the direct pQCD production cross section at NNLO
is sensitive to $M_t$.
This provides us with an important conclusion:  if TeV scale black holes
are indeed formed at the LHC, then one signature of this
will be an unusually copious production of massive (Top quarks, Higgs and
SUSY) particles, which
is not possible via pQCD processes. Hence if we observe
very high rates for massive particle production at the LHC, this might provide
indirect evidence that TeV scale black holes are being produced.

The paper is organized as follows.  In Sec. II we present the
computation for the rate of top quark production
from black holes via Hawking radiation at the LHC.
In Sec. III we sketch the pQCD techniques at NNLO
for top quark production at the LHC.
In Sec. IV we present and discuss our results.

\section{Top Quark production from black holes at the LHC }

If black holes are formed at the LHC then they will quickly evaporate
by emitting thermal Hawking radiation. The emission rate per unit time
for top quark with momentum $p =|\vec p|$ and
energy $Q= \sqrt{p^2+M_t^2}$ can be written \cite{gram} as
\be
\frac{dN}{dt }=
\frac{c_s \sigma_s}{8\pi^2}\frac{dp\, p^2}{(e^{Q/T_{BH}} + 1)}\,,
\label{thermal}
\ee
where $\sigma_s$ is the grey body factor and $T_{BH}$ is the black hole
temperature, which depends on the number of extra dimensions and on
the TeV scale Planck mass. $c_s$ is the multiplicity factor.
The temperature of the black hole is given in \cite{pp}, namely
\be
T_{BH}=\frac{d+1}{4\pi R_{S}} ~=~
\frac{d+1}{4 \sqrt{\pi}}~
M_P [\frac{M_P}{M_{BH}}\frac{d+2}{8\Gamma(\frac{d+3}{2})}]^{\frac{1}{1+d}}\,,
\ee
where $R_{S}$ is the Schwarzschild radius of the black hole,
$M_P$ is the TeV scale Planck mass, $M_{BH}$ is the mass of the black hole
and $d$ is the number of extra dimensions. 
The grey body factor in the geometrical approximation is
given by \cite{gray4,bv,grayd}
\be
\sigma_s = \Gamma_s 4\pi (\frac{d+3}{2})^{2/(d+1)} \frac{d+3}{d+1} R^2_{S}\,,
\label{grey}
\ee
where we take $\Gamma_s=\frac{2}{3}$  for spin half particles.
The total number of top quarks emitted from the black holes is thus given by:
\be N_{\rm top ~quark}= \int_0^{t_f} dt \int_0^{M_{BH}} dp\,
\frac{c_s \sigma_s}{8\pi^2}\frac{p^2 }{(e^{\sqrt{p^2+M_t^2}/T_{BH}} + 1)}\,,
\label{NN}
\ee
where $t_f$ is the total time taken by the black hole to completely evaporate,
which takes the form \cite{pp1}:
\be
t_f~=~\frac{C}{M_P}(\frac{M_{BH}}{M_P})^{\frac{d+3}{d+1}}\,.
\ee
$C$ depends on the extra dimensions and on the polarization degrees
of freedom, etc. However, the complete determination of $t_f$
depends on the energy density present outside the black hole
which is computed in \cite{cooper} where
the absorption of the quark-gluon plasma \cite{plasma1}
by a TeV scale black hole at the LHC is considered
(this time is typically about $10^{-27}$ sec).
The value we use throughout this paper is $t_f$= $10^{-3}$ fm
which is the inverse of the TeV scale energy.

This result in Eq. ({\ref{NN}}) is for top quark emission from
black holes of temperature $T_{BH}$.
To obtain the top quark production cross section from all
black holes produced in proton-proton collisions at the LHC
we need to multiply the black hole
production cross section with the number of top quarks produced from a
single black hole.
The black hole production cross section
$\sigma_{BH} $ in high energy hadronic collisions at zero
impact parameter is given in \cite{pp,cham}, namely
\bea
\sigma_{BH}^{AB \rightarrow BH +X}(M_{BH})
= {\sum}_{ab}~
\int_{\tau}^1 dx_a \int_{\tau/x_a}^1 dx_b f_{a/A}(x_a, \mu^2)
\nonumber \\
\times f_{b/B}(x_b, \mu^2)
\hat{\sigma}^{ab \rightarrow BH }(\hat s) ~\delta(x_a x_b -M_{BH}^2/s).
\label{bkt}
\eea
In this expression $x_a (x_b)$ is the longitudinal momentum
fraction of the parton inside
the hadron A(B) and $\tau=M_{BH}^2/s$, where $\sqrt s$ is the
hadronic center-of-mass energy.
Energy-momentum conservation implies $\hat s =x_ax_b s=M_{BH}^2$.
We use $\mu = M_{BH}$ as the scale at which the
parton distribution functions are measured. ${\sum}_{ab}$ represents
the sum over all partonic contributions.
The black hole production cross section in a binary partonic
collision is given by \cite{pp}
\bea
\hat{\sigma}^{ab \rightarrow BH }(\hat s) =
\frac{1}{M^2_P} [\frac{M_{BH}}{M_P}
(\frac{8\Gamma(\frac{d+3}{2})}{d+2})]^{2/(d+1)}\,,
\label{bk3}
\eea
where $d$ denotes the number of extra spatial dimensions.
The total cross section for top quark production at LHC is then given by
\be
\sigma_{\rm top ~quark} = N_{\rm top ~quark} \sigma_{BH}.
\label{susybk}
\ee
We will compare this cross section for top quark
production via black hole resonances with the top quark cross section
produced via pQCD processes at NNLO, as will be explained in the next
section.

\section{ Top Quark Production via pQCD Processes at the LHC}

The top quarks at LHC are mainly produced in $t\bar t$ pairs.
At the LHC proton-proton collider, the QCD production process
involves quark-antiquark and gluon-gluon fusion mechanism.
The gluon-gluon fusion processes give the dominant cross section
(about 90 percent). This subprocess at high energy is the main reason
for larger rate of the cross section compared to Tevatron at Fermilab.
The single top quark production occurs via electroweak process. The single
top quark production cross section ($\sim $ 300 pb ) is smaller
compared to $t\bar t$ total cross section ($\sim 970$ pb) at LHC at
$\sqrt{s}$ =14 TeV pp collisions. Hence we will not consider the single top quark production cross section
\cite{sing} in this paper. We will consider $t \bar t$ pair production using
parton fusion processes at LHC and will
compare them with the top quark production cross section from black holes.

At the next-to-next-to-leading order (NNLO) one needs to compute the
following partonic subprocesses. On the leading-order (LO) level we have
\bea
q + {\bar q} \rightarrow t {\bar t},~~~~~~~~~~~g + g \rightarrow t {\bar t}.
\label{LO}
\eea
In NLO we have in addition to the one-loop virtual corrections to the above
reaction the following two-to-three body processes
\bea
q + {\bar q} \rightarrow t {\bar t}+g,~~~~~~~~~~~g+q ({\bar q}) \rightarrow t {\bar t}+q({\bar q}),
~~~~~~~~~~~g + g \rightarrow t {\bar t}+g.
\label{NLO}
\eea
At NNLO level we receive the two-loop virtual corrections to the LO processes in eq. (\ref{LO})
and one-loop virtual corrections to NLO reactions in eq. (\ref{NLO}). To these contribution
one has to add the results obtained from the following two-to-four body reactions
\bea
&& g + g \rightarrow t {\bar t}+g+g, ~~~~~~~~~~~~~~~g+g \rightarrow t{\bar t}+q+ {\bar q}, \nonumber \\
&& g+q ({\bar q}) \rightarrow t {\bar t}+q({\bar q})+g, \nonumber \\
&& q + {\bar q} \rightarrow t {\bar t}+g+g, ~~~~~~~~~~~~~~~q+{\bar q} \rightarrow t{\bar t}+q+ {\bar q}, \nonumber \\
&& q +  q \rightarrow t {\bar t}+q+q, ~~~~~~~~~~~~~~~{\bar q}+{\bar q} \rightarrow t{\bar t}+{\bar q}+ {\bar q}, \nonumber \\
&& q_1 +  q_2 \rightarrow t {\bar t}+q_1+q_2, ~~~~~~~~~~~~~~~q_1+ {\bar q}_2 \rightarrow t{\bar t}+q_1+  {\bar q}_2. \label{NNLO}
\eea
After the phase space integrals has been done the partonic cross section ${\hat \sigma}$ is
rendered finite by coupling constant renormalization, operator renormalization and the removal of collinear divergences. The renormalization scale $\mu_R$ is set equal to the mass
factorization scale $\mu_F$. The cross section
for top quark production in proton-proton collisions at the LHC
is given by
\begin{eqnarray}
\label{eqn4.1}
\sigma=\sum_{a,b=q,\bar q,g} \int dx_1 \int dx_2\,f_a(x_1,\mu_F^2)
f_b (x_2,\mu_F^2 ) ~{\hat \sigma}_{ab}
\end{eqnarray}
where ${\hat \sigma}_{ab}$ is the partonic level cross section for top quark production.
For the details, see \cite{toplhc1,vogt}. Reviews of present status of top quark physics
at LHC can be found in \cite{toplhc2}.

\section{Results and Discussions}

In this section we will compute the top quark production
cross section from black hole at $\sqrt{s}$ = 14 TeV in pp collisions
and will compare them with the top quark production via parton fusion
processes at NNLO. The top quark production from black holes
is described in section II. For the black hole production
we choose the factorization and normalization scale to be
the mass of the black hole. As the temperature of the black hole at the LHC
is $\sim$1 TeV there is not much difference in the top quark
production cross section from black holes if the top quark mass $M_t$
is increased from 165 to 180 GeV. For black hole mass $M_{BH}$ much
closer to the Planck mass $M_P$ the string corrections are important.
In this situation string ball production becomes important \cite{cham}.
For this reason we will choose black hole mass $M_{BH}$ to be larger than
the Planck mass $M_P$ \cite{cham,cooper,susy,higgs} in our computations below.

\begin{figure}[htb]
\vspace{2pt}
\centering{\rotatebox{270}{\epsfig{figure=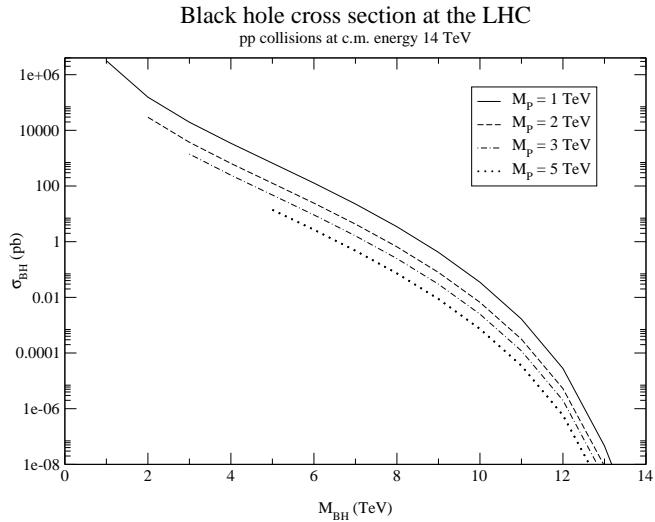,height=10cm}}}
\caption{ Total cross sections for black hole production at the LHC.}
\label{fig1}
\end{figure}

In Fig. 1 we present the black hole production cross section
at the LHC.  The $y$-axis
is the black hole production cross section $\sigma_{BH}$ in
pb and the $x$-axis is the black hole mass $M_{BH}$ in TeV.
The solid, dashed, dot-dashed and dotted curves are for
Planck masses of 1, 2, 3 and 5 TeV respectively.
The number of extra dimensions $d=4$. As can be
seen from the figure the cross sections decrease rapidly when
both the Planck and black hole masses increase. These
black hole production cross sections will be
multiplied with the number of top quarks produced from a
single black hole to obtain the top quark production cross section
from a black hole at the LHC.

\begin{figure}[htb]
\vspace{2pt}
\centering{{\epsfig{figure=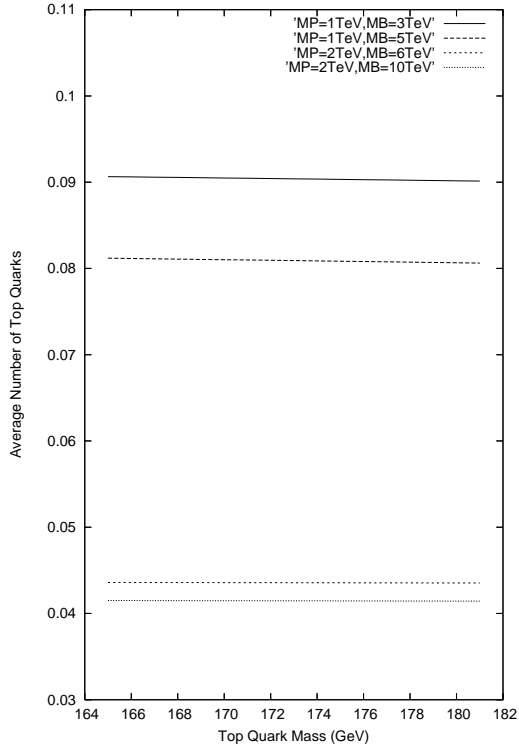,height=10cm}}}
\caption{ Average Number of top quark production from a single black hole
at LHC. The upper two lines are for black hole masses equal to 3 and
5 TeV respectively with the Planck mass equal to 1 TeV in
each case. The lower two lines are for black hole masses equal to 6 and
10 TeV respectively with the Planck mass equal to 2 TeV in
each case.}
\label{fig2}
\end{figure}

In Fig. 2 we present results for the average number of top quarks produced
from a single black hole as a function of top quark mass.
The $y$-axis is the average number of top quark production
from a single black hole and the $x$-axis is the mass of the top quark in GeV.
The upper two lines are for black hole masses equal to 3 and
5 TeV respectively with the Planck mass equal to 1 TeV in
each case. The lower two lines are for black hole masses equal to 6 and
10 TeV respectively with the Planck mass equal to 2 TeV in
each case.It is clear that the average number of top quark produced
from a single black hole is much larger for smaller black hole mass. This is
because as the mass of the black hole becomes smaller the temperature
becomes larger ($\sim$ TeV) and the thermal radiation of top quarks
are enhanced. This is the case from a single black hole emission.
The black hole production cross section itself decreases at LHC as the
mass of the black hole increases. Hence the total cross section
of top quark production from black holes at LHC is a competitive effect
from the above two factors (see eq. (\ref{susybk})).

\begin{figure}[htb]
\vspace{2pt}
\centering{{\epsfig{figure=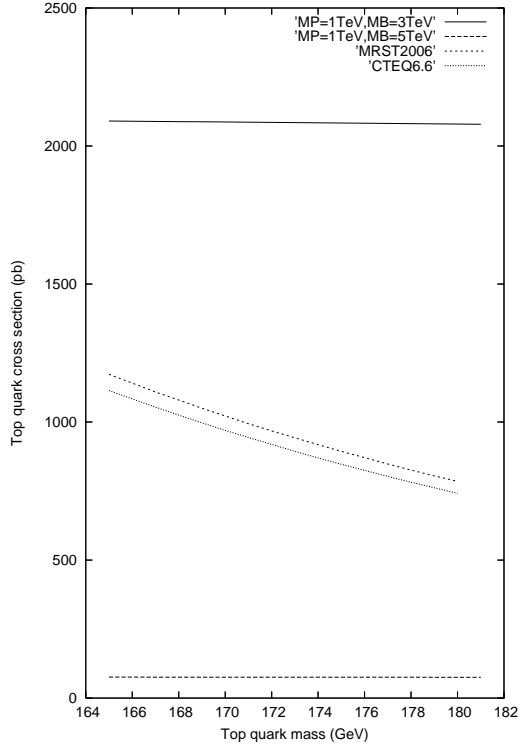,height=10cm}}}
\caption{ Total cross section for top quark production at LHC
from black holes and from direct pQCD processes at NNLO.
The two middle curves are NNLO results and the upper and lower
curves are from black holes of masses 3 TeV and 5 TeV respectively with
the Planck mass equals 1 TeV in each case.}
\label{fig3}
\end{figure}

In Fig.3 we present the total top quark production cross section
from black hole production and compare them with the pQCD predictions at NNLO.
The former is given for Planck mass $M_P=1$ TeV
and for two different choices of the black hole mass, namely
$M_{BH}= 3,5 $ TeV respectively. We plot for comparison the NNLO
top quark cross section from \cite{vogt} with $\mu_F=\mu_R=M_t$.
The two middle curves are NNLO results and the upper and lower
curves are from black holes. The upper NNLO curve is for MRST 2006
PDF and the lower NNLO curve is for CTEQ6.6 PDF. The upper black hole
curve is for black hole mass equal to 3 TeV and the lower black hole
curve is for black hole mass equal to 5 TeV with
the Planck mass being 1 TeV in both the cases. For larger Planck mass
the cross section becomes even smaller and hence we do not plot them.
It is clear that the total cross
section via black hole production is larger than the pQCD cross section
for small $M_P$ and $M_{BH}$ and is not sensitive to the increase in top
quark mass.

In summary, we have computed top quark production cross section from black holes
in proton-proton collisions at the LHC at $\sqrt{s}$ = 14 TeV
via Hawking radiation within the model
of TeV scale gravity and have compared it with the pQCD cross sections at NNLO.
As the temperature of the black hole is $\sim$ 1 TeV there is a huge amount of top quark
production from black holes at the LHC if the Planck mass is $\sim$ 1 TeV and the black hole
mass is $\sim$ 3 TeV. We also find that, unlike standard model predictions,
the top quark production from black hole is not sensitive to the increase in top quark mass.
Hence we suggest that the measurement of an increase in cross section for
heavy particle (top quark or Higgs \cite{higgs} or SUSY \cite{susy}) production at the LHC
can be a useful signature for black hole production.

We make a brief comment about the grey body factor used in this paper.
The grey body factor which is used in Eq. (\ref{grey}) is only valid in the regime of massless quanta and when the energy of the emitted particle is small compared to the black hole mass. Therefore, the computed cross section 
 eq. (\ref{susybk}) gives only an approximation to the actual cross section. 
 Since this approximation improves for more massive black holes, our conclusions should remain valid. Finally, we warn the reader that the Planck mass MP = 1 TeV used in this
paper is somewhat at odds with the constraints posed by LEP data on contact
interactions [34] which suggests the Planck mass is greater than 2.2 TeV. If
these constraints are true, then black hole production at the LHC, albeit
exciting, may not lead to measurable contributions even if the large extra
dimension scenario is realized.

\acknowledgments

We would like to thank Jo Ashbourn for her continuing encouragement for Andrew's last paper
to be completed and published. The work of G.C.N. was supported in part by the U.S. Department of
Energy under Grant no. DE-FG02-01ER41195.

\end{document}